# Traverse the landscape of the mind by walking: an exploration of a new brainstorming practice

Xiaofeng Wang, Daniel Graziotin, Juha Rikkilä, and Pekka Abrahamsson
Faculty of Computer Science
Free University of Bozen-Bolzano, Bolzano/Bozen, Italy
{xiaofeng.wang, daniel.graziotin, juha.rikkila, pekka.abrahamsson}@unibz.it

**Abstract:** Group brainstorming is a well-known idea generation technique, which plays a key role in software development processes. Despite this, the relevant literature has had little to offer in advancing our understanding of the effectiveness of group brainstorming sessions. In this paper we present a research-in-progress on brainstorming while walking, which is a practice built upon the relationship between thinking and walking. The objective is to better understand how to conduct group brainstorming effectively. We compared two brainstorming sessions, one performed during a mountain walk, the other traditionally in a room. Three preliminary findings are obtained: walking can lead to an effective idea generation session; brainstorming while walking can encourage team members to participate in and contribute to the session in an equal manner; and it can help a team to maintain sustainable mental energy. Our study opens up an avenue for future exploration of effective group brainstorming practices.

# INTRODUCTION

The early phase of software development has been considered fuzzy ideation, a miraculous moment of sudden apparition which is nearly impossible to explain (Risku, 2010). Ideas seem to come out of nowhere and they look to be generated by blessed people. Yet it is important for software companies to have a better understanding of their idea generation processes, since the ability to generate good ideas is considered a characteristic of successful business ventures (McFadzean, 2000). Ideas and ideation processes are some of the most important assets for a company's future product creation. If well performed, ideation can deliver information on the concept and meaning of a product, its positioning in product portfolio, and the value estimation of the product for the users and the company (Hellström & Hellström, 2002).

The process of idea generation is becoming less mysterious as research progresses, even though it is still challenging (Kultima, 2010). As idea generation processes are likely to be structured, the enhancement of creative processes looks feasible and becomes an important subject of study. However, popular software and system development methods lack techniques and tools to support ideation processes. They are commonly used after the ideation and suitable approaches when "problem is known, solution is unknown". But, they are inadequate when "problem is unknown" (Blank, 2008), which are typical issues to be tackled in ideation processes.

Brainstorming is arguably the most used idea generation technique. It is often used in the informal and early phase of software design (e.g., (Wu, Graham, & Smith, 2003)) and has been incorporated into software development processes as a standard idea generation technique (Paetsch, Eberlein, & Maurer, 2003). In software development, requirements are often elicited from the brainstorming sessions of developers and stakeholders (Soundararajan & Arthur, 2009). Since software development is a collaborative process usually undertaken by one or more development teams, brainstorming on software design is conducted by a group of people. However, the research on brainstorming in Psychology studies reveals that group brainstorming generally suffers from productivity loss due to various factors. Even though there are suggested rules to follow (M. Diehl & Stroebe, 1987; Michael Diehl & Stroebe, 1991; Mullen & Johnson, 1991), the research on group brainstorming, specifically in software development, lacks proposals of alternative ways to deal with the reported inefficiency and effectively enhance the process.

Our study aspires to find effective ways to conduct group brainstorming and proposes group brainstorming in a context different from traditional meeting rooms: brainstorming while walking. It is based on the relationship between thinking and walking that can be spotted both in scientific studies (e.g., Neurology) and in humanistic essays (e.g., Philosophy). We conducted a pilot study to show how it is different from brainstorming in rooms. As a first step, we obtained several preliminary findings regarding brainstorming while walking that can be further investigated.

The rest of the paper is organized as follows. In the next section, we report the body of knowledge on group brainstorming and present the scientific and philosophical root of brainstorming while walking. Section 3 contains the research approach that explains how we conducted the pilot study, which will serve as the basis for the full study eventually. The results from the pilot study are presented in Section 4. The last section carries the results further and summarizes three preliminary findings regarding brainstorming while walking and concludes the paper with the outlook of future work.

# BRAINSTORMING AND WALKING

## Group Brainstorming

Brainstorming, originally introduced by Osborn (1953), is arguably the most popular idea generation technique. Although it was born as a specific tool for creative idea generation, with a specific set of rules to be followed (Osborn, 1953), nowadays the process is rather non systematized and is becoming an umbrella term for group ideation process (Kultima, 2010).

Defined in the Oxford dictionary as "a way of making a group of people all think about something at the same time, often in order to solve a problem or to create good ideas", brainstorming is usually conducted in groups. However, the studies in Psychology have consistently discovered productivity loss during group brainstorming processes and suggest that they are not as effective as brainstorming separately as individuals (Mullen & Johnson, 1991).

There are various explanations why productivity loss happens in group brainstorming sessions. Production blocking is one reason (M. Diehl & Stroebe, 1987). In brainstorming only one person may gainfully voice his or her ideas in a group at any given time. Meanwhile, individuals may forget or suppress their ideas because they seem less relevant or less original at a later time. Additionally, being forced to listen to ideas of other people may be distractive and interfere with an individual's own thinking (M. Diehl & Stroebe, 1987). The productivity loss is largely due to the delay between idea generation and verbalization which might cause participants to suppress their ideas or forget them (Michael Diehl & Stroebe, 1991).

Evaluation apprehension can be another reason for productivity loss. The fear of negative evaluations from other group members prevents people who are working in groups from presenting their more original ideas. Social inhibition would be greater if more group members perceive other members as experts (M. Diehl & Stroebe, 1987).

Yet another reason for productivity loss is free riding, because the identifiability of individual contributions is low in group brainstorming sessions, as well as the perceived effectiveness of individual contributions. It is claimed that if each group member is expected to receive an equal share of benefits, the highest benefit an individual can obtain is when contributing the least, due to self-interested laziness (M. Diehl & Stroebe, 1987).

In spite of productivity loss, group brainstorming in collaborative work such as software development is inevitable, especially in the context of design activities (Shih, Venolia, & Olson, 2011; Wu et al., 2003) and requirements engineering (Nuseibeh & Easterbrook, 2000; Paetsch et al., 2003). Several rules for effective group brainstorming, including focus on quantity, withhold criticism, welcome unusual ideas and combine and improve ideas, have been proposed in the original brainstorming literature (Osborn, 1953). It is claimed that, if these rules are followed, a group of individuals could produce better results in terms of the quantity and quality of the ideas produced. However, it is not clear how these rules have been or can be effectively used in brainstorming activities of software development teams due to the lack of studies.

Apart from focusing on group dynamics, there are no studies (as far as the authors are aware of) of conducting brainstorming in different physical formats other than traditional meeting. Some variation has been introduced in practice. For example, instead of sitting around a table in a room, group members stand up in a circle and face to each other while speaking. Some physical movements such as passing or throwing balls between team members are adopted to facilitate the conversation and flow of discussion. One purpose of the standing ups and physical movement is to energize people (Lightmsith & Lightsmith, 2012).

## Walking and Thinking

Research in Neurology has accumulated knowledge on the relationship between physical activity and brain function. There is evidence that physical activity is linked with improvements in brain function and cognition (Hillman, Erickson, & Kramer, 2008). Among different activities, the profound impact of walking on thinking has been explored by various authors at various times. According to the author of "Wanderlust: A History of Walking" (Solnit, 2001), walking is the state in which the mind, the body and the world are aligned. It is believed that once the walking rhythm has been established, one becomes much more alert to minute variations in sensory input (such as smell, colour or temperature). Meanwhile, the mind starts to wander much more freely. As the author put it eloquently: "this creates an odd consonance between internal and external passage, one that suggests that the mind is also a landscape of sorts and that walking is one way to traverse it" (Solnit, 2001).

Quite a few famous historical figures representing creativity and profound thoughts are linked to walking. For example, walking was Wordsworth's means of composition. Nietzsche turned to solitary walks for recreation. Rousseau portrayed walking as both an exercise of simplicity and an opportunity for contemplation. In his "Reveries of the Solitary Walker", he said that "these hours of solitude and meditation are the only ones in the day during which I am fully myself and for myself, without diversion, without obstacle, and during which I can truly claim to be what nature willed" (Rousseau, 1980). Danish philosopher Kierkegaard claimed that a lone walker was both present and detached, and the mind worked best when surrounded by distraction (Solnit, 2001).

However, Thoreau in his "Walking" questioned of thinking anything else other than walking itself. "I am alarmed when it happens that I have walked a mile into the woods bodily, without getting there in spirit. […] What business have I in the woods, if I am thinking of something out of the woods?" Yet even Thoreau himself admitted that "[…] the thought of some work will run in my head and I am not where my body is - I am out of my senses." (Thoreau, 1862)

Combining walking and thinking is seen as one of the best ways to generate ideas and get things moving in practice (Avalon, 2010). Walking has been recommended as an "enabler" for idea generation. It is considered "creative loafing time" (Al., Thompson, Evison, & Podmoroff, 2012), which allows the mind to have some rest, and it will often come up with connections precisely when it is not trying to make them. Walking also presents a change of environment. Sometimes changing the setting changes the thought process (Al. et al., 2012). There is some evidence from game development industry that walking, especially out in the natural environment, is a helpful activity for coming up with good ideas (Kultima, 2010).

In spite of all these recommendations, there is very little empirical evidence to show how useful walking is as a physical format of group brainstorming.

## RESEARCH METHODOLOGY

Our study intends to be the first exploration of brainstorming while walking; therefore the primary objective of our study is to provide an initial understanding of the phenomenon. The main research methodology adopted in our study is case study, which is an approach used to study phenomena in their natural contexts (Yin, 2003). The unit of analysis is a group. A brainstorming session conducted by a group is treated as a case. We adopt a comparative case study strategy and compare brainstorming while walking with more traditional manners, such as brainstorming in the room.

We conducted two pilot cases, one group doing brainstorming while walking, the other in the room. The two cases were chosen following a convenience sampling method. We studied the brainstorming sessions of two groups of university students majored in Computer Science at our university. The two groups of students were from the same course. As part of the course, they were developing games for entrepreneur education. The groups were formed at the beginning of the course based on self-organization. Table 1 shows the profiles of the two teams, namely the All-Ins team and the Raising Legacy team.

| Table 1. The Profiles Of The Two Teams | | | | |
|---|---|---|---|---|
| Team Name | No. of Students | Ave. Age | Ave. Years of Study | Brainstorming Style |
| The All-Ins | 4 | 20 | 1 | Outdoor walking |
| Raising Legacy | 5 | 27 | 2 | Indoor meeting |

An iterative approach was adopted by both teams to organize their game development processes over the course duration. The two brainstorming sessions were held at the early stage of the game designing when they had to brainstorm on what types of games they would develop and the initial elements of the games.

The only derivation of our research approach from the standard non-participative case study approach is that we intervened with one group, asked them to do the brainstorming while walking before the session started, and provided one piece of guidance during the session. No intervention has been introduced to the other group who followed their normal approach and brainstormed in the room. We classify our research approach as action case study since we did introduce some actions in one case in order to better understand the research topic (Braa & Vidgen, 1999). Action case is a term coined in (Braa & Vidgen, 1999) to refer to a hybrid of interpretation and intervention. In such a method the degree of intervention is much lower than that in a pure action research study. In our case, our main purpose and emphasis is on understanding not changing the phenomenon. Our research method is also different from experiment study since we have not controlled the setups of the two sessions, the topics to brainstorm on, or the processes they should follow.

The All-Ins team was walking on the mountain while brainstorming. The session lasted about 150 minutes. The Raising Legacy team remained in the classroom for the whole session. The session lasted about 180 minutes. Both sessions were video recorded. A brief retrospective session was held with each team after the session to understand their feelings and impressions of their respective sessions. Each team was asked to draw a mood chart to illustrate their feelings as a group during the brainstorming session using the Niko-cale symbols (also known as Niko-niko calendar, as shown in Table 2) which makes visible as for the mood of a team and the feelings, motivation and morale of the team (Sakata, 2006). In addition, a brief group retrospective was held with the All-Ins team to obtain their opinions of brainstorming while walking.

| Table 2. The Niko-cale Symbols | |
|---|---|
| Face Mark | Meaning |
|  | Pleasant, Happy or Good |
|  | Ordinary |
|  | Unpleasant, Unhappy, Not Good or Bad |

The videos were analysed using the temporal bracketing strategy, one approach to analyse process data (Langley, 1999). Each video was decomposed into a set of successive "periods". Table 3 and Table 4 show the high level temporal blocks of the two sessions. These periods do not have any particular theoretical significance. They are a way of structuring the description of the sessions, and enable the constitution of comparative units of analysis of them. With this strategy, a shapeless mass of process data is transformed into a series of more discrete but connected blocks. Our main analysis focus is the team walking on the mountain. The team staying in the room is treated as a comparison case to better understand the brainstorming process of the mountain group.

## PRELIMINARY FINDINGS

A comparative analysis of the two pilot cases revealed several key differences between the two brainstorming sessions. In this section, we report each case from the three aspects where the key differences were manifested: time spent on group discussion, group interactions, and team mood change during the brainstorming session.

An IRB approval for conducting empirical studies on human participants is not required by our institution. Informal consent was obtained from all participants.

### Brainstorming while Walking

The All-Ins team walked on a mountain trail near the university site. It is a trail with beautiful views and mixed woods. The route was chosen by the team themselves. During the session the team walked about 8 kilometres and the altitude change was about 200 meters. Table 3 shows the periods of the team's brainstorming while walking session.

| Table 3. The brainstorming while walking session of the All-Ins team | |
|---|---|
| Time Duration | Activity |
| 15 minutes | Walking from the university building to the starting point of the mountain trail. No observable group discussion. Casual chatting. |
| 40 minutes | Walking up hill. No observable group discussion. Casual chatting. |
| 10 minutes | Walking in flat. Group discussion started. |
| 22 minutes | Stopped walking. Intensive group discussion. |
| 18 minutes | Mixed walking and breaks. Group discussion continued. |
| 45 minutes | Walking down and back to the university. No observable group discussion. Casual chatting. |

The All-Ins team started walking from the university site. It took them 15 minutes to reach the starting point of the trail. The team chose to start walking from the steeper side of the trail. The steep slope turned out to be tough for the team members to walk, especially for the girl in the team. There were no observable group discussions on the game design happening during the first 55 minutes. Some casual chats happened spontaneously. During the interview, the team members commented that they were not able to think and talk when they were walking up because: "*We were stressing out on the hill*", "*during the walk it's difficult to think about the game because you are concentrating on the walk*". However, they claimed that this time was not completely wasted: "*We were thinking [about the game] during the walking*".

When they reached the flat part they started group discussion. At the given stage, they felt they had to stop and catch breath. Here we suggested that they stop walking and concentrate on brainstorming. This was followed by more than 20 minutes of really intensive and active group discussion on their game design. This was the highlight of the whole session. The discussion among the team members was lively, interactive and energetic. The group interactions among the team members during this period were active and even. There was not an obvious centre of communication. Everyone was attentive and contributing.

No one was left out. The team members have talked with one another constantly. Yet they were not fighting with each other to express their own opinions. The discussion flew naturally and smoothly from one member to the next. The interactions seem effortless and enjoyable. All the team members concentrated on the task. No casual chatting happened. One team member recalled what happened during this period: *"[during the break] the ideas, like, came suddenly. Maybe the walking thing, like, helped us"*

The team resumed walking after this intensive discussion period, while continuing to discuss the design issues. But the discussion died out naturally when the focus was back on walking again. After almost another 20 minutes the team decided to walk back to the university. During the walking down the hill, no observable group discussion happened. There were casual chats again on other topics not relevant to their game design.

In total, the actual time the team spent on group discussion is around 30% of the whole session. They described how the whole process worked: *"When we started, we didn't have a really good idea of what we were gonna do. While we were walking, we were kind of thinking about it, and then when we stopped, we got the things on paper"*. At the end of the brainstorming session, the team members said that they felt "good", "relaxed". Figure 1 is the mood chart the team created that reflected the team mood during the whole session.

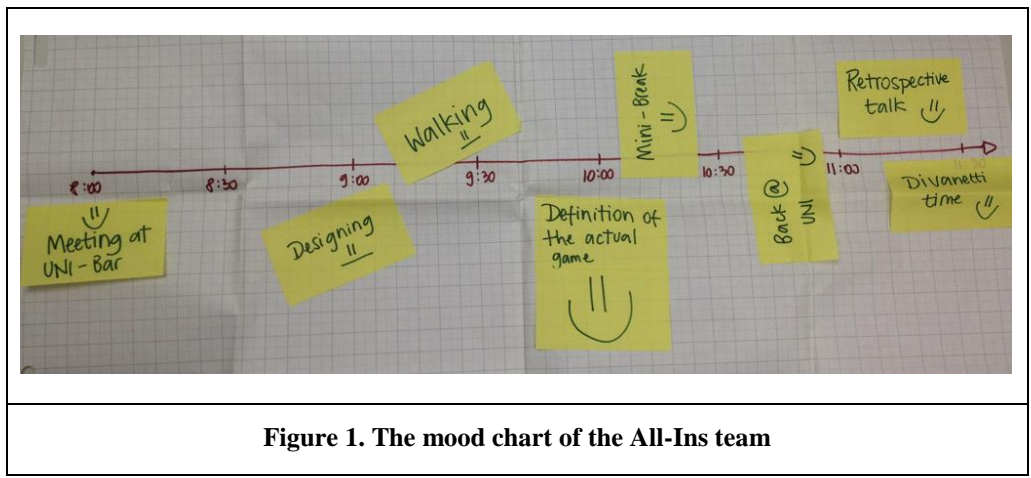

**Figure 1. The mood chart of the All-Ins team**

It can be seen that the mood of the team has been kept positive and the spirit is especially high during the second half of the session. The team attributed the somehow flat mood in the first half of the session to the steep hill they had to walk on which stressed out the team. The mood chart is consistent with what we observed from the video recording of the team session. After the initial physical fatigue, the team members kept smiling and joking while discussing their game design.

## Brainstorming in the Meeting Room

The Raising Legacy team spent about 180 minutes sitting in a classroom to brainstorm their game design. The classroom was playing the role of a typical large sized meeting room, which can be found in many companies. Table 4 shows the periods of the team's brainstorming in room session. Except for a coffee break in the middle, most of the time the team was discussing the game design in group or researching on the related issues individually. Apart from some casual chatting, the overall time the team spent on group brainstorming was more than 70% of the whole session. The casual chatting happened more often in the second half of the session, which brought out the laughter from the team members and lightened the atmosphere. However, the team always went back to the design issues right after the chatting.

| Table 4. The brainstorming while walking session of the All-Ins team ||
|---|---|
| Time Duration | Activity |
| 20 minutes | Started with 4 team members. Sitting in the room and discussing in group. No casual chatting. |
| 70 minutes | The 5th member joined the team. Discussing in group mixed with researching individually. Some casual chatting. |
| 15 minutes | Break. Left the room for coffee in group. |
| 5 minutes | Work resumed. The 5th member who came late left. |
| 66 minutes | Continued with 4 team members. Discussing in group. More casual chatting. |

The team formed two sub-groups almost right from the beginning. There was very little interaction between the two sub-groups. One team played a bridging role between the two sub-groups and therefore became the centre of the communication. Except for him all the other team members remained at their own seats throughout the whole session. It seems that the team was confined by the emerging sub-groups and the table that separated them. Cross sub-group discussions happened, but very rarely.

Figure 2 shows the mood chart produced by the team at the end of the session. It can be seen that the mood of the team had quite some ups and downs in comparison to the mood chart of the All-Ins team. In addition to the subjective mood chart, the mood peaks can be spotted from the video recording as well. The team members looked more active and cheerful near the middle of the video recording. Their body language was more expressive and their movements increased. They discussed intensively. However, their excitement rapidly decreased before and after the coffee break.

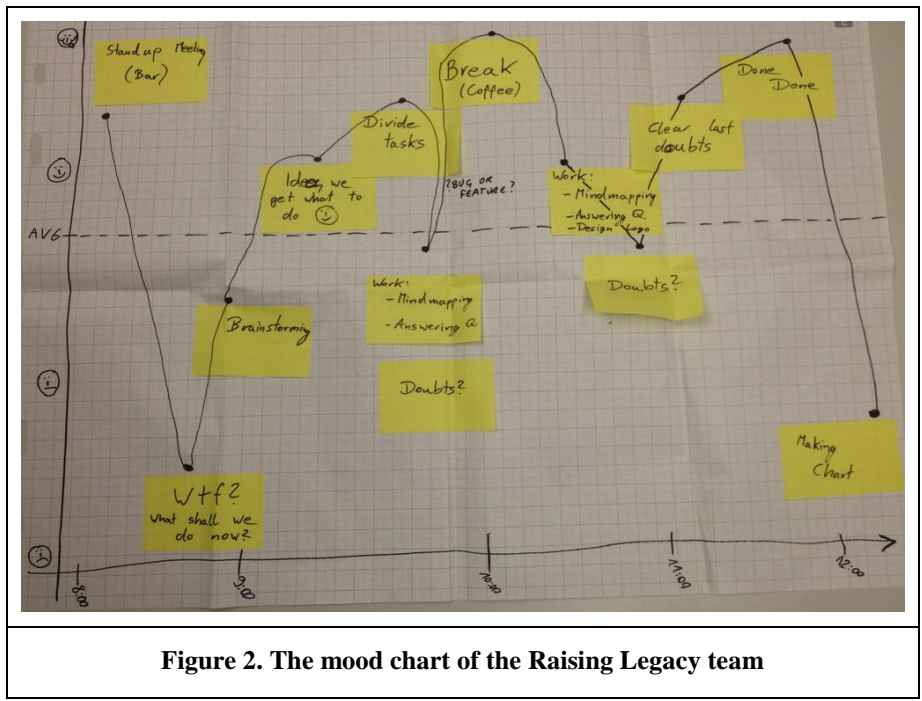

**Figure 2. The mood chart of the Raising Legacy team**

# DISCUSSION AND FUTURE WORK

The two cases form a sharp comparison in terms of the actual amount of time spent on group discussion: around 30% in the case of the team walking on the mountain, and more than 70% in the case of the team sitting in the room. It seems that a lot of time was wasted in the first case, even though the mountain team claimed that they were thinking individually about the game design while walking.

It has been claimed that the most effective brainstorming technique would let individuals first think alone and generate ideas; then to let the group evaluate the generated ideas (Isaksen, 1998). When producing creativity in groups, it is often the case that initial ideas are created in solitude (Kultima, 2010). A problem solver may need to spend significant amounts of time thinking alone, thus apparently resulting in poor group idea generation performance (Skowronski, 2004). The case of the mountain team shows that walking may help naturally reconcile the stress on continuous communication among people, and allow individuals to think alone and let the group evaluate the ideas afterwards.

It took a long time for the mountain team to start meaningful group discussion in comparison to the room team which started group discussion almost immediately after the session began. The delay may be due to the fact that the mountain team chose to start from the steeper side of the hill and were stressed out on the hill in the beginning. If the walk were less demanding physically, the team may have entered the group discussion stage earlier. Actually, it is suggested that the mind works the best at "three miles per hour" walking pace (Solnit, 2001). The mountain team case shows that the walking should not be too demanding to enable the thinking. In spite of the demanding walk however, an effective brainstorming session did happen afterwards – still during the outing session. The team did experience a sort of "serendipity" moment during the intensive brainstorming session.

The two teams brainstorming in two different settings show different group dynamics in terms of interactions. The group interactions of the mountain team were more intensive, interactive and lively. The team members all participated in the discussion evenly and contributed to it equally. No free riding happened. The attitudes of the team members can be described as positive. In comparison, the group dynamics of the room team were not the same. Not all the team members contributed equally. The interactions were pivoted on one team member who played the central communication role. Almost no interactions happened between the two sub-groups formed by the division of the table. The team members sat at the same place throughout the session except for the person who was bridging the communication between the two sub-groups. In comparison, no physical object existed in between the mountain team members while they were walking. There was no fixed walking pattern either in terms of who is next to whom while walking. The open and free space among the team members and changing physical vicinity may have prevented any sub-group formation and encouraged free communication among them.

The fluctuation of team mood during a brainstorming session can impact the results of the session. Mood and feelings are representation and reflection of mental energy (Kurtzberg & Amabile, 2001), and they are deeply linked to creativity processes such as idea generation and evaluation according to the research in Psychology and Cognitive Science (Baas, De Dreu, & Nijstad, 2008; Davis, 2009). Therefore, mental energy is an important factor for brainstorming. Figures 1 and 2 show that both the All-Ins and Raising Legacy started their brainstorming sessions with good moods. However, the fluctuation patterns of the team moods are quite different. It is interesting to notice that the mood chart of the mountain team showed a more stable, smooth and mostly positive mental feeling during the whole session, in spite of the physical challenges presented during the brainstorming session up and down the hill. On the contrary, while the room team had much less physical movement, the team's mood had gone through quite a few ups and downs as shown by their mood chart. It can be argued that the mountain team maintained a good mood and therefore more sustainable mental energy throughout the session than the room team did.

In the future we will replicate the pilot study in a more rigorous manner, ideally in industrial settings and address the potential validity threats in the pilot study. We will also investigate the link between brainstorming while walking and the creativity or quality of the ideas generated afterwards, to quantify the benefits that brainstorming while walking can produce. Another interesting direction to explore is how to utilize different tools that support mobility and collaborative work in brainstorming while walking sessions to improve the productivity of group brainstorming in open air, e.g., to record generated ideas, or to provide supporting materials for idea generation in an easily accessible and user friendly manner.

## ACKNOWLEDGEMENTS

We thank the two groups of students who gave us consents to take the videos of them and collaborated on this study.